# When gravity meets philosophy again: the "Gravitas project"

**Matteo Tuveri** [a,b,*], **Daniela Fadda** [c], **Viviana Fanti** [a,b] **and Walter Bonivento** [b]

[a] Physics Department, University of Cagliari, Complesso Universitario di Monserrato, 09042, Monserrato, Italy,

[b] INFN Sezione di Cagliari, Complesso Universitario di Monserrato, 09042, Monserrato, Italy

[c] Department of Pedagogy, Psicology, and Philosophy University of Cagliari, via is Mirrionis, Cagliari, Italy

*E-mail:* matteo.tuveri@ca.infn.it, danielafadda@unica.it, viviana.fanti@ca.infn.it, walter.bonivento@ca.infn.it

Gravity is, by far, one of the scientific themes that have most piqued the curiosity of scientists and philosophers over the centuries. From Aristotle to Einstein, from Hawking to now on, scientists' have always put a creative effort to solve the main puzzles of the understanding of our universe: why do things move, the birth of the cosmos, dark matter, and dark energy are just a few examples of gravity-related problems. Philosophers are interested in this field, too, and when they have met physicists' needs, a new conceptual revolution has started. However, since Einstein's relativistic theories and the subsequent advent of quantum mechanics, physicists and philosophers have taken different paths, both kidnapped by the intrinsic conceptual and mathematical difficulties inherited by their studies. A question arises: is it possible to restore a unitary vision of knowledge, overcoming the scientific-humanistic dichotomy that has established itself over time? The answer is certainly not trivial, but we can start from school to experience a new vision of a unified knowledge. From this need, the "Gravitas" project has been born. "Gravitas" is a multidisciplinary outreach and educational program devoted to high school students (17-19 years old) that mixes contemporary physics and the philosophy of science. Coordinated by the Cagliari Section of the National Institute of Nuclear Physics, in Italy, "Gravitas" has started on December 2021 with an unconventional online format: two researchers coming from different fields of research (physics vs philosophy, history of science, scientific communication) meet a moderator and informally discuss about gravity and related phenomena. The public can chat and indirectly interact with them during the YouTube live using Mentimeter. The project involves about 250 students from 16 high schools in Sardinia, Italy. Students have also been involved in the creation of posts thought for social media platforms whose content is based on the seminars they attended during the project. We present the project and discuss its possible outcomings concerning the introduction of a multidisciplinary approach in teaching physics, philosophy, and the history of contemporary physics in high schools.

---

[*]Speaker

**1.     Introduction**

Physical and philosophical (conceptual) questions drive human investigations of nature across the centuries: why and how do things move? how is the universe made of? The answer is a matter of gravity. From Aristotle, to Einstein, passing through Galilei and Newton, gravity is a concept that we thought we had mastered since the origin of the human being but continues to perplex us, considering recent developments in High Energy Physics and Astrophysics. The universe as it appears to us today is still full of mysteries, of concepts that we struggle to comprehend, including dark matter, dark energy, inflation, black holes, and their origins. These concepts are under-appreciated and are given minimal attention and explanation in secondary, undergraduate, and master level of university education and yet have strong philosophical implications and societal implications [1-4]. After 20th century scientific revolutions (general relativity and quantum mechanics) physics and philosophy have taken different paths. Is it possible to restore a unitary scientific vision? The answer is certainly not trivial, but we can start from schools to offer a new way to approach to science, where there are no boundaries between knowledge but, instead, everything is linked as it was, and it is in the process of learning how the nature is made [5-6]. From this need, the "Gravitas" project has been born.

"Gravitas" is a multidisciplinary outreach and educational program devoted to high school students (17-19 years old) that mixes contemporary physics and the philosophy of science. The project is coordinated by the Cagliari division of the National Institute for Nuclear Physics (INFN) along the outreach activities of the DARK collaboration. Its main goal is to recombine the conceptual and practical nature of physics offering a multidisciplinary (contemporary) vision of science and knowledge. For this reason, the scientific committee was made by more than 50 physicists and philosophers of Italian and European Universities and Institutions with the goal to discuss and decide the scientific content of the project. Indeed, the themes dealt belong to $20^{th}$ century physics and philosophy of science, history of science, science communication and logics.

Activities were divided in two parts. In the first part (from December '21 to April '22), we organized a series of 16 online seminars (due to pandemic) called "Nuovi Dialoghi sui Massimi Sistemi" – "New Dialogues Concerning the (Two Chief) World Systems", dealing with different topics concerning physics, history of science, philosophy, and science communication. In the second part (from April '22 to September '22), the project was devoted to the organization of the first edition of the "GravitasFest", a festival held in Cagliari on the $17^{th}$ and $18^{th}$ of September. We will not discuss the festival here, but we just mention its innovative bottom-up organization: themes, formats, speakers were chosen together with students who attended the first part of the project, who were also the protagonists of the voluntary team and of an art exhibition made by their own work as science communicators.

As researchers, we were also interested in monitoring some peculiar aspects of informal learning of science mediating by the project and its activities. We were interested in monitoring students' motivation towards physics and philosophy on specific items and on how the project can influence students' perception on physics and philosophy, science communication and students' interest in these three fields [7-9].

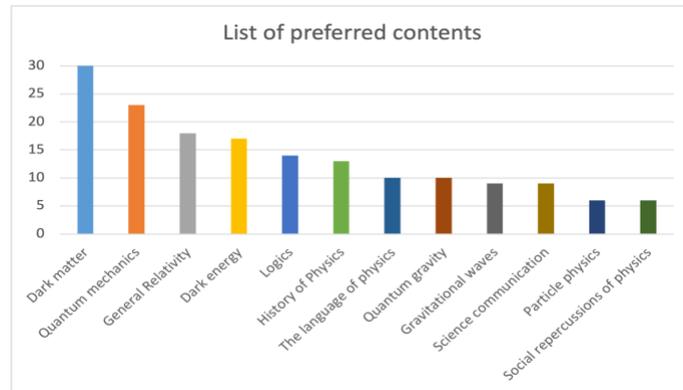

Fig. 1 List of arguments discussed during the online seminars with experts. The panel shows the number of preferences for each content.

## 2.     Methods

In the first part of the "Gravitas" project, we collected the participation of 32 scientists from European Universities and Institutions who gave their scientific contribution on various topics, from the birth of quantum mechanics and general relativity to black holes and gravitational waves, passing through quantum gravity, the language of physics, the concept of matter from the 18$^{th}$ century to nowadays, the social impact of science, dark matter, and dark energy. Online seminars were organized with an innovative format: two researchers coming from different fields of research (physics vs philosophy, history of science, scientific communication) meet a moderator and informally discuss about gravity and related phenomena. The public can chat and indirectly interact with them during the YouTube live using Mentimeter. Videos were transmitted and stay located at the INFN Cagliari Youtube channel.

Concerning students attending the online activities, 236 students (m=128, f=108) from 16 high schools in Sardinia (43 were 19 yo; 130 were 18 yo; 63 were 17 yo). Only 127 students (m=72, f=55; 111 from scientific, 9 from "humanities", 7 from artistic high schools) completely ended the project (writing the posts as discussed in the following). As a mandatory activity to conclude the project, students created some material such as posts, one post for each month (4 posts in total on 4 different arguments) aimed for the socials (such as Facebook and Instagram). Topics related to their work come from what they learnt from the seminars. They could also work in group. The post had a fixed structure: it should contain information about the name and surname of the authors, their school and class, one image, and a caption. They should also include some references (for the images and for the information included in the text) and some hashtags. In this way, they had the possibility to manage the learning process in such a way they act as science communicators, preparing materials to diffuse science on a specific platform and for a specific target using that platform. Fig. 4 shows the list of preferred contents to write a post according to the students. The material has been used to organize an art-exhibition to also enhance the work made by students. Thus, we transform the material sent by them in qr-codes and we manage them to create some constellation visible in the month of September 2022, when the festival was held (see later).

Concerning research, also following arguments in Refs [7-9], we wrote a research questionnaire to investigate the influence of the project on students' interest and motivation towards physics, philosophy, and science communication.

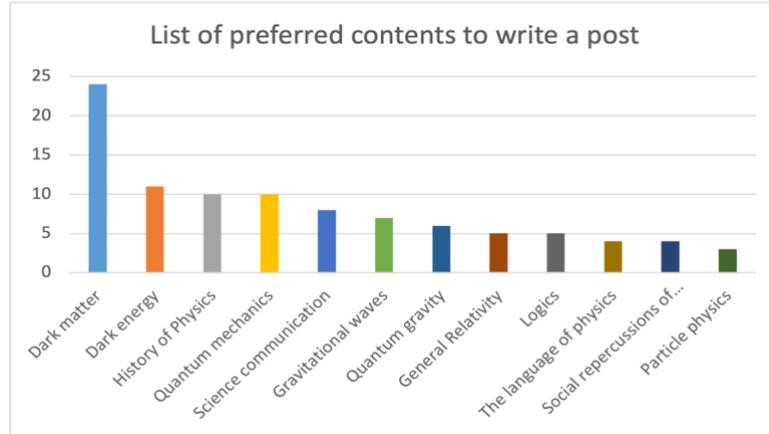

Fig. 2 List of preferred contents to write a post. The panel shows the number of preferences for each content. Note that data arise from the questionnaire, not from the number of posts we collect.

We also investigated students' feelings about a possible implementation of Gravitas methodology in schools. We leave the analysis of the latter for future publications. Students could answer by using a 6-points Likert scale, from 1 (completely disagree) to 6 (completely agree). We collect 70 answers (m=42, f=28) in the period April (end of the project) – June 2022 and qualitatively analyzed them by calculating means and related standard deviation. To determine whether there is any statistically significant difference between the means of the items based on gender, we carried out an analysis of variance (ANOVA, not shown here). For each investigated topic, the Cronbach's alpha is higher than 0.8. We also wrote a questionnaire devoted to teachers who attended the project with their classes, but we leave the discussion for a future paper [10].

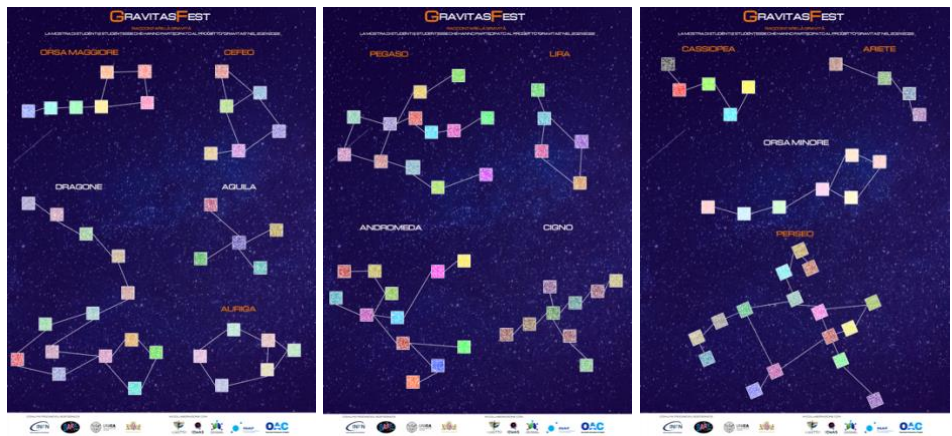

Fig. 5 The artwork with qr-codes about students' posts. On the left, Ursa Mayor, Cepheus, Dragon, Eagle, and Auriga constellations. On the center, Pegasus, Lyra, Andromeda, Cygnus constellations. On the right, Cassiopeia, Aries, Ursa Minor and Perseus constellations. On the bottom of the three charts, the list of organizers, partners, and collaborators of the festival.

## 3.    Results

Concerning some data about visualization of the online seminar, in the period December 2021 - July 2022, the average number of visualizations per YouTube video is around 270 (80 students on average during the live sessions). As outcome of the project related to students' work inspired by the content of the seminar, there have been the art-exhibition. Indeed, the posts have been transformed from a simple exercise of science communication to a form of collective artwork as shown in Fig. 5.

Concerning the questionnaire, questions and related results are shown in Figs 6, 7a and 7b.

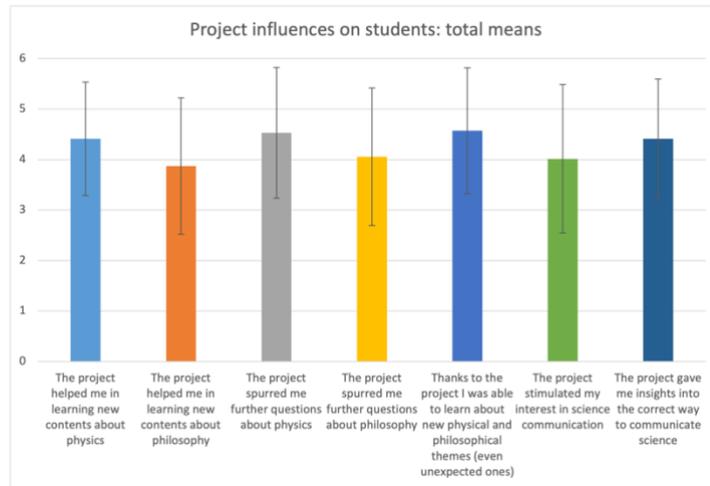

Fig. 6 The panel shows the means related to the influence the project has had on students about physics, philosophy, and science communication. Error bars are the standard deviation.

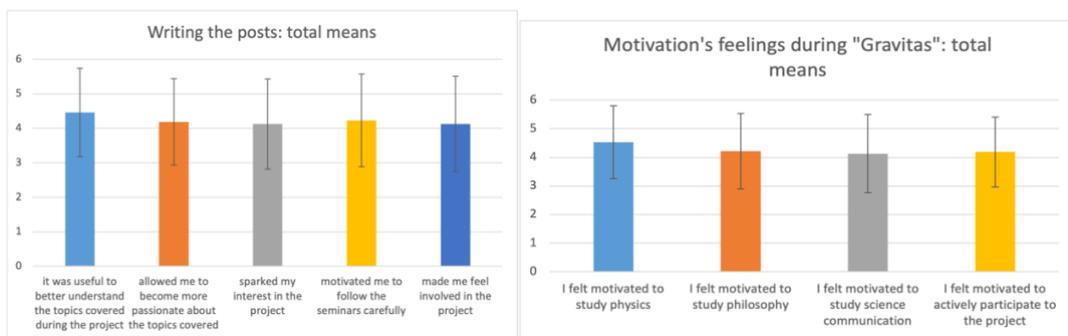

Fig. 7. On the left (a), the panel shows means related to students' feeling on writing the posts. On the right (b), students' motivation in specific topics while they were attending the project. Error bars are the standard deviation. The ranking scale goes from 1 (completely disagree) to 6 (completely agree).

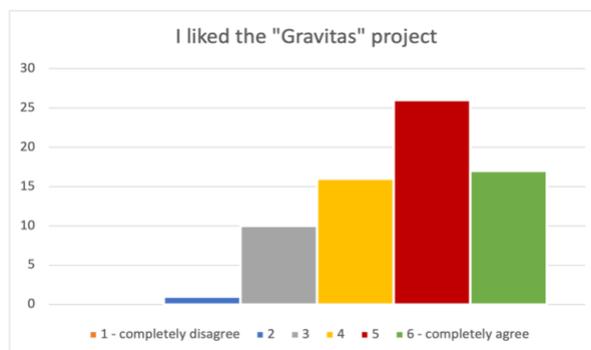

Fig. 8. Students' overall perception of the project. The ranking scale goes from 1 (completely disagree) to 6 (completely agree)

## 4. Discussion and conclusion

Participation of students during the first part of the project was high and satisfactory.

The interest in this kind of activities is confirmed by the average number of visualizations per video in 6 months which is good considering that the videos are located on an institutional channel of INFN Cagliari, where the rate of activities is not high (difficult to reach if you are not invited to go there). Qualitative analysis based on questionnaire's results shows an overall good influence on students' attitude towards physics, philosophy, and science communication (means higher than 4), with higher means in physics with respect to the others. They appreciated experiencing the Gravitas' format questionnaire, they appreciated writing posts and considered the activity as useful to better understand the topics covered during the project. They also felt motivated in actively participate to the project and its activities.

The work has some limitations: the sample is small, so we cannot infer something general about the positive outcomes we registered in the first edition of the project. For this reason, we would like to extend the audience to all Italy in next years. In conclusion, the project had a positive influence on students' motivation towards physics, philosophy, and science communication (means related to physics are higher than the others). They felt motivated in actively participate to the project and its activities. Finally, when asked to report their feeling about the overall projects, students really liked the project (Fig. 8).